\documentclass[twoside]{article}

\usepackage[sc]{mathpazo}
\usepackage[T1]{fontenc}
\usepackage{microtype}
\usepackage{amsmath}
\usepackage{mathtools}
\usepackage{amssymb}
\usepackage{graphicx}
\usepackage[hmarginratio=1:1,top=32mm,columnsep=20pt]{geometry}
\usepackage[font=it]{caption}
\usepackage{subcaption}
\usepackage{paralist}
\usepackage{multicol}
\usepackage{mathtools, cuted}

\usepackage{lettrine}

\usepackage{abstract}

\usepackage{titlesec}
\renewcommand\thesection{\Roman{section}}
\titleformat{\section}[block]{\large\scshape\centering}{\thesection.}{1em}{}

\usepackage{fancyhdr}
\usepackage{hyperref}

\title{\vspace{-15mm}%
	\fontsize{24pt}{10pt}\selectfont
	\textbf{Decision Making on Fitness Landscapes}
	}	
\author{%
	\large
	\textsc{R. Arthur${}^{*1}$ and P. Sibani${}^{2\sharp}$} \\[2mm]
	\normalsize	{\it ${}^1$University of Exeter, Wellcome Trust Centre for Biomedical Modelling and Analysis. }\\
	\normalsize	{\it ${}^2$Department of Physics, Chemistry and Pharmacy, University of Southern Denmark. } \\
	\normalsize	${}^*$\href{mailto:rudy.d.arthur@gmail.com}{rudy.d.arthur@gmail.com} ${}^\sharp$\href{mailto:paolo.sibani@sdu.dk}{paolo.sibani@sdu.dk}
	\vspace{-5mm}
	}
\date{}

\begin{document}

\maketitle
\thispagestyle{fancy}

\begin{abstract}
\noindent 
We discuss fitness landscapes and how they can be modified to account for co-evolution. We are interested
in using the landscape as a way to model rational decision making in a toy economic system. We develop
a model very similar to the Tangled Nature Model of Christensen et. al. that we call the
Tangled Decision Model. This is a natural setting for our discussion
of co-evolutionary fitness landscapes. We use a Monte Carlo step to simulate decision making and
investigate two different decision making procedures.
\end{abstract}

\begin{multicols}{2}
\section{Introduction}

\lettrine[nindent=0em,lines=3]{S}ewall Wright's fitness landscape \cite{Wright:1932} 
posits that an individual's reproductive success is a function of its genes.
Under evolutionary pressure the genomes of a population shift to a 
peak on the fitness landscape where they  remain until a higher fitness peak is discovered
by a random mutation.

The idea has been used in theoretical evolutionary biology,
e.g. in Kauffmann's NK model~\cite{Kauffman:1989}. 
The same model has been used by Levinthal \cite{Levinthal:1997}, Rivkin \cite{Rivkin:2000} and 
numerous others to study human organizations. To mention some of the early and highly cited work: 
Kauffmann and collaborators worked on the idea of a technological landscape
\cite{Auserwald:2000,Kauffman:2000};  Ethiraj and Levinthal worked on organizational design landscapes
\cite{Ethiraj:2004a, Ethiraj:2004b}; Gavetti and Levinthal modelled cognitive search processes on landscapes \cite{Gavetti:2000}.

The fitness landscape metaphor is  powerful but difficult in many respects. The  valleys 
assumed to separate fitness peaks may not exist in reality, peaks may be connected by 
selectively neutral `ridges' \cite{Gavrilets:1997}. Visualizing high dimensional landscapes is hard and 
counter-intuitive~\cite{Arrival:2014}. However, the main issue for us is that 
the Malthusian fitness of an individual, intended as its ability
to reproduce,    should depend not only on the the individual's   own genes
 but also on the populations of the other agents making up the  eco-system.
 This is the view point taken in the Tangled Nature Model (TNM) of 
evolution~\cite{Christensen:2002}, which has been studied extensively 
as a model of biological ecologies  \cite{Laird:2006, Laird:2007, Laird:2006a} 
and which more recently has been extended to study systems of human organizations \cite{Jensen:2013, Nicholson:2016, us:2016}.
In these later works, evolution is by mutation and
selection, as  in  biological systems.
 In this work,  the TNM is enlarged by  introducing agents which
 observe the dynamics  while  actively making decisions 
regarding the fate of new mutants generated by
 reproductive  errors. The resulting model, which we dub
 TDM for \emph{Tangled Decision Model}, offers a new 
 approach to discuss decision making based on faulty and/or
 incomplete information in an interacting, multi-species environment.

Consider the environment of a lemur. There are non-biological 
aspects like rivers and climate but one is forced to mention trees providing fruit 
and habitat, insects as prey and the fossa as a predator. These other species have environments 
whose description includes lemurs. Thus their fitness landscapes are intimately intertwined and 
changes by one species imply changing fitness values of the others along the genetic co-ordinates 
describing their mutual interaction e.g. as prey become better at evading predators the predator's fitness decreases. 
A changing landscape means a changing 
selective difference between nearby points on the fitness landscape and new evolutionary paths can open as a result.

This notion of co-evolution is simply that species evolve together. 
This entails  a continuous fluctuation and rearrangement of their fitness landscapes 
in response to the changes in the fitness and population of others. The same is clearly true when 
taking a sociological or economic view. Companies exist in a market with other companies, 
whose   products they consume  to make their own, which in turn are sold to  other firms or to  the general public.
A change in the inputs or outputs of one company  can clearly affect both suppliers and consumers.
Consider refining ore. This industry uses the products of the mining and energy industries to produce
metals   which are sold to e.g. auto manufacturers, soft drink companies or jewellers.
Development of a more efficient extraction technique  or of a new lightweight alloy will have
consequences for and change the `fitness' of all the companies which supply or use products of a refinery.

Simple examples make clear that a one-to-one mapping of species, industry, organization or agent to a single fitness 
value that is constant for all time is not realistic. Such a static landscape picture will only apply in the 
very restricted setting of one species in a constant environment or as an approximation
for  a very short timescale. In the traditional fitness landscape
different species' landscapes are unconnected and
unaffected by the sizes of each other's populations. 
The real fitness landscape seen by a 
single species should be continually shifting due to interactions with other species (as well as potential changes 
in the non-biological, regulatory or consumer environment).
 Evolution, either by rational agents attempting to increase 
their fitness or by blindly generated  biological mutants, can
 move the species uphill, but can also cause the hill 
itself to shrink. By incorporating co-evolution the fitness landscape as usually envisioned 
is a metaphor  stretched to its  breaking point.

The dynamic nature of fitness landscapes has been explored before see e.g. \cite{Mustonen:2009} where the term 
`fitness seascape' is used. This is a useful perspective but, in order to calm the seascape,
we move up a level of description, from the  landscape of individuals to a landscape of eco-systems.
In this paper we construct a `system fitness landscape'
where every point represents an eco-system whose fitness
is the sum of the fitnesses of all extant individuals. 
The higher this quantity, the more growth will occur and the system is said to be more `fit'.
There is no competition or co-operation between different systems rather,
as the individuals in  a system 
grow, evolve, compete and co-operate,
the system performs a walk in the system fitness landscape.

As mentioned, our TDM  is in some respects  identical to the well-studied Tangled Nature Model (TNM) of biological
evolution~\cite{Christensen:2002}. The difference is the  
introduction of boundedly rational agents that 
try to optimise the fitness of their species by making `decisions', while observing the model's dynamics.
The usual TNM agents will be dubbed `irrational'.
We introduce two explicit decision making processes, local and global, and
parameters controlling the degree of rationality. We then investigate co-evolution of both irrational 
and rational agents and how different decision making processes affect global optimization.

\section{Fitness Functions}\label{sec:fitnessfunctions}
Let a species be specified 
by a genome $g$ consisting of $L$ binary variables. When we talk about species 
we have in mind both biological and economic species. In the latter,
`species' are organizations or products as spelled out in section \ref{sec:model}. The `population'
is to be thought of as size under some metric like number of employees or
market share.

Consider first a single non-interacting species. A fitness function $f(g)$ takes genomes and returns 
real numbers. We then have another function 
$P$ taking fitness values and returning the probability , $0< P(f(g))<1$, for that species to reproduce or grow during
a certain time interval. We require $P$ to be monotonic so that $f(a) < f(b) \implies P(f(a)) < P(f(b))$. 
$P$ should also be a function of the population of the species
, $N$, since there must be physical limits to growth. This implies that $P$ 
should be a sigmoid (s-shaped) function such as
\begin{equation}
P(g,N) = \frac{1}{ 1 + \exp(- f(g) + \mu(g) N ) }.
\end{equation}
This form is not unique but serves our purposes. $\mu$ is a damping term that fixes the carrying capacity, smaller
$\mu$ means higher total populations are possible. 

Now consider a system of two species, labelled $1$ and $2$ with populations $N_1$ and $N_2$. 
Species $1$ has a fitness $f(g_1;g_2,N_1,N_2)$ depending on the other species present as well as its 
population. $P$ has the same form as before
\begin{align}
&P(g_1;g_2,N_1,N_2) = \\ \nonumber &\frac{1}{ 1 + \exp( - f(g_1;g_2,N_1,N_2) + \mu N ) } \nonumber
\end{align}
with $N = N_1 + N_2$ and 
we have assumed that $\mu$ is constant across species so they all use the same amount of `resources'. 

Note that the fitness, $f$, can depend on the population of the extant species. 
A system with nine wolves and one rabbit 
should not be fit with any reasonable definition of fitness! Size affects fitness, a fact that is missing in the usual 
fitness landscape. To clarify this we split the fitness of an individual into a part depending 
only on the species itself and a part depending on its interaction with the other species:
\begin{equation}
f(g_1;g_2,N_1,N_2) = V(g_1) + E(g_1;g_2,N_1,N_2).
\end{equation}
The total population is controlled by $\mu$ which can be thought of as a physical constraint like
the size of a lake or, with our economic landscape, available employees or the total amount of demand. We assume that
interactions $E$ only depend on population ratios instead of absolute populations. This means
if the physical constraint is decreased by moving to a bigger lake or opening a new
market the different species do not become
more or less fit relative to one another. This may not be the case in practice, e.g. if one species is
much faster at multiplying, but we assume it as a first approximation.
We will also assume that $E$ depends linearly on the population ratio 
$$
E(g_1;g_2,N_1,N_2) = G(g_1;g_2) n_2 
$$
where $n_2 = \frac{N_2}{N}$. 
This lets us rewrite
\begin{align} \nonumber
&P(g_1;g_2,N_1,N_2) = \\ 
&\frac{1}{ 1 + \exp( - V(g_1) -G(g_1;g_2) n_2 + \mu N ) }
\end{align}
 
We now straightforwardly extend this to $S$ species. For species $a$ the fitness is
\begin{align} \nonumber
&f_a(g_a; g_1,\ldots,g_S,n_1, \ldots, n_S) = V(g_a) + \\
&E(g_a; g_1,\ldots, g_S,n_1, \dots n_S).
\end{align}
We assume linear dependence on the population ratios and that $E$ can be broken down into
a sum over pairwise interactions. This means we can describe the interaction of all other species with species
$a$ using an edge-weighted, directed graph instead of the more general situation which would require 
a hypergraph \cite{Sonntag:2004}. These assumptions amount to
\begin{align} \nonumber
E(g_a; g_1,\ldots,g_S,n_1, \ldots, n_S) &= \sum_i^S G(g_a; g_i) n_i \\
&= \sum_i^S J_{ai} n_i
\end{align}
where
$$
J_{ai} = G(g_a; g_i) \neq J_{ia} {\rm for } \quad i\neq a \quad {\rm and } J_{aa} = 0.
$$
We define the fitness of a single individual of species $a$ by
\begin{equation}\label{eqn:singF}
f_a = V(g_a) + \sum_i^S J_{ai} n_i
\end{equation}
the fitness of the whole species $a$ by 
\begin{equation}
F_a = N_a f_a
\end{equation}
and the fitness of the whole system by 
\begin{equation}\label{eqn:totF}
F = \sum_a^S F_a
\end{equation}
 
If we assume that the interaction independent part of the fitness, $V$, is zero we have a
model of co-evolution only. With an appropriate implementation of mutation and death processes this model is 
the Tangled Nature Model (TNM)\cite{Christensen:2002}. We have 
chosen the functional form of $P$ arbitrarily, used the same value of $\mu$ for all species and
made assumptions about the dependence of an individual's fitness on the number and type of other species.
Nevertheless this argument shows that the TNM models co-evolution on interconnected fitness landscapes. 
 
Usually fitness landscapes are defined over `genetic' space. By this we mean the function
$f_a(g_a; g_1,\ldots,g_S,n_1, \ldots, n_S)$ is thought of as a function of the $L$ binary variables
$g_a$ only. This fitness landscape is a real
valued function defined over an $L$ dimensional hypercube. Because this approach suppresses the other
variables the landscape will shift unpredictably in response to changes in the populations of the other species. 
A less chaotic picture is obtained by thinking about the system landscape.
The total number of species is $S = 2^L$, though most of the time 
most of them will have zero population. We rewrite the global fitness function as a sum over all $2^L$ species
\begin{equation*}
F(N_1, \ldots, N_{2^L}) = \sum_a^{2^L} N_a \left( V(g_a) + \frac{ \sum_i^{2^L} J_{ai} N_i }{ \sum_i^{2^L} N_i } \right).
\end{equation*}
The system fitness landscape is a function, $F$, defined over a $2^L$ dimensional lattice with positive integer co-ordinates 
$(N_1, \ldots, N_{2^L})$.

\section{Rational Decision Making}
The work of Levinthal \cite{Levinthal:1997} and others using fitness landscapes generated by the NK model 
to explore rational descision making is interesting to re-examine from this new point of view. The
NK model is a method to make a `tunably rugged' fitness landscape, in our language it is a way to
choose $V$. One typically investigates
adaptive walks on this landscape, with multiple agents starting at different points.
These agents do not interact directly but those with higher fitness survive longer and multiply more
until only agents at local maxima remain. This has been an important testing ground for questions about epistasis, 
the inter-dependence of traits, however
recent work in sociology and economics has focussed on dynamic modelling of coupled systems.\footnote{ Mostly using differential
equation models, see e.g. Minsky https://sourceforge.net/projects/minsky/ .} 

A rational agent on an NK fitness landscape who can see a short distance, i.e. evaluate
the fitness difference between its current and potential positions, will choose to move towards the 
configuration with the highest fitness. If there is death and reproduction, species with higher fitnesses survive for longer
until the species with highest fitness outcompetes all the others. Mutation allows exploration of the landscape,
but eventually the system will find itself with no accessible higher fitness peaks leading to equilibration at a local maximum. 
In this picture the only ways an agent would move is if its vision was increased 
so it could see other, higher peaks that were previously invisible. Kauffmann used the term `long-jump' for the
process where a new genome is chosen randomly and selected if the fitness is higher.
Alternatively, the agent could make mistakes with a certain probability and travel downhill
where higher peaks may be visible: this can be important on a very rugged fitness landscape. 
In practice forecasting can improve, lucky mistakes occur and there are completely
unexpected innovations. However the NK-model has all species on the same landscape and 
does not consider the effect that each move has on the landscape itself.

We can take a Schumpeterian view that `creative destruction'
periodically rearranges the landscape and makes movement possible. If we are trying to model
whole systems then this view is unsatisfactory - the shocks should be generated by the system
itself not exogenously. The landscape should change when a decision made by one firm affects the 
landscape of all others, these firms then react and change the landscape of the original mover. 
Kauffmann introduced co-evolution to the NK model \cite{Johnsen:1989}
but we have shown that co-evolutionary landscapes can be modelled very naturally in the framework of
the TNM, to which we now turn.

\subsection{A Simple Model of Decision Making}\label{sec:model}
We describe a model of rational searches on fitness landscapes using a Monte Carlo approach:
the Tangled Decision Model. In this model we 
have products with a blueprint given by the genome, $g$. These could be multiple products
produced by a single company or different companies which manufacture one product.
Each company/product (or the people behind it) attempts to grow as much as possible within the environmental
constraints. As we have emphasised, changes in one company/product/species affect the others. One way to simulate this is
to allow a small amount of evolution on a frozen landscape, update the landscape, perform another small evolution step,
etc. The TNM uses the reproduction and death of discrete agents to accomplish this, as will our model. We do not
think of a company as composed of many independent actors, the reproductive agents 
are simply a mathematical artifice that enable a simple simulation framework, similar to
discretising a differential equation. The true actors in the model are whole species. For more discussion
about this point see \cite{us:2016}.

Our setup is similar to the TNM (see \ref{sec:app} for more details),
except for the way mutants are treated. First
we assign the values of the matrix $J_{ab}$ from a symmetric random distribution, there is
no correlation between the values of $J_{ab}$ and $J_{a'b}$ for similar $a$ and $a'$; in the NK language
the landscape is very rugged. Such correlations can be included, \cite{Andersen:2016}, but we do not do so in this work.
We start with a small population of one random species. The update step is: 
\begin{compactitem}
\item choose an agent with uniform probability and delete it with probability $p_{kill}$
\item if the agent survives, compute $p_a = \frac{1}{1 + e^{-f_a + \mu N}}$ and
reproduce with that probability. 
\end{compactitem}
Repeating these two steps $N/p_{kill}$ times counts as one `generation'. This a natural timescale for the model
independent of $N$. We show results in terms of generations.

We assume that every company tries to innovate by giving up a small amount of its growth potential to
generate new products/genomes. Choosing which of these innovations to fund should be a rational decision made
by CEOs, venture capitalists or investment banks. We use a Monte Carlo step to represent
the decision makers' choice. When an agent is chosen
to reproduce it copies itself but with a probability $p_{mut}$ to
flip each genome `bit'. This mutation process turns the genome $g_a$ into $g_b$. 
We evaluate the fitness of $a$, $f_a$ and of $b$, $f_b$ and compute
\begin{equation}
p^l_{ab} =
	\begin{cases} 
      m \exp(-\beta(f_a - f_b) ) & f_b < f_a \\
      m & f_a = f_b \\
      1 - (1 - m) \exp(-\beta(f_b - f_a) ) & f_b > f_a 
   \end{cases}
\end{equation}
for some constants $\beta,m$, with $0<\beta<\infty$ and $0<m<1$. We produce the new species
with that probability, otherwise we reproduce the original. Using $\beta$ we can interpolate between a completely 
rational regime where no fitness reducing steps are allowed, $\beta = \infty$, or a completely irrational regime where 
any change is accepted, $\beta = 0$, as in the standard TNM. The functional
form  of $p^l_{ab}$ is sigmoid and equals $m$ when $f_a = f_b$. Standard Monte
Carlo has $m = 1$, so fitness increasing steps are always accepted. Here we use $m = \frac{1}{2}$, this makes it
equally likely to reject moves that increase fitness and to accept decreases. We refer to this process as
`local' decision making.

We can also consider the global point of view. We produce a mutant $b$ in the same manner and add it to the system. 
We compute the fitness of the {\it system} including the new mutant $F^{+b}$. We then compute the fitness 
of the system with one extra member of species $a$, $F^{+a}$. We add $b$ with probability
\newline\newline
\resizebox{.5\textwidth}{!}
{
$
p^g_{ab} =
	\begin{cases} 
      m \exp(-\gamma(F^{+a} - F^{+b}) ) & F^{+b} < F^{+a} \\
      m & F^{+b} = F^{+a} \\
      1 - (1 - m) \exp(-\gamma(F^{+b} - F^{+a}) ) & F^{+b} > F^{+a}
   \end{cases}
$
}\newline\newline
otherwise we add $a$. $0 < \gamma < \infty$ is used to interpolate between perfect and imperfect
descisions and $m = \frac{1}{2}$ again. We will refer to this process as `global' decision making. 
This kind of decision making is possible 
in a command economy - an innovation will not be funded if it undermines other key industries. Even a regular investor
might not proceed with a product if it would conflict with his other investments. From the single
company point of view, where all the products are sold by the same parent corporation, the global
perspective is even more natural, companies don't want to undermine themselves!

Local and global can be combined by computing
\begin{equation}\label{eqn:mc}
p_{ab} = p^l_{ab} p^g_{ab}.
\end{equation}
Global optimisation can be thought of as exploring the system landscape and local
optimisation as exploring the individual species landscape. For low values of $\gamma$ and $\beta$
the decision makers often choose badly. If we think that decision makers would not willfully choose a bad
product or reject a good one it is equivalent to think of $\gamma$ and $\beta$ as controlling the
accuracy of the information that the decision makers recieve.
The functional form chosen makes larger differences in fitness less likely to be mistaken.

Since interactions are uncorrelated we are effectively saying that the new products which appear have no 
relation to the other extant products. In reality much innovation is steady improvement of existing goods. 
By introducing correlated interactions we could, by changing the degree of
correlation, interpolate between incremental and drastic innovation in the same way that changing the K parameter in the 
NK model creates smooth or rugged landscapes.
It is known from previous work on the TNM that correlations do not lead to qualitatively new behaviour \cite{Andersen:2016} so
for simplicity we have only investigated the uncorrelated case.

\section{Results}
In this section we will focus on key illustrative results and basic phenomonology rather than 
an exhaustive sweep of parameter space. The values $\beta = 1, \gamma=1$, correspond to quite
low Monte Carlo `temperatures' and it is quite rare to make fitness lowering moves.
\includegraphics[angle=-90,width=0.45\textwidth]{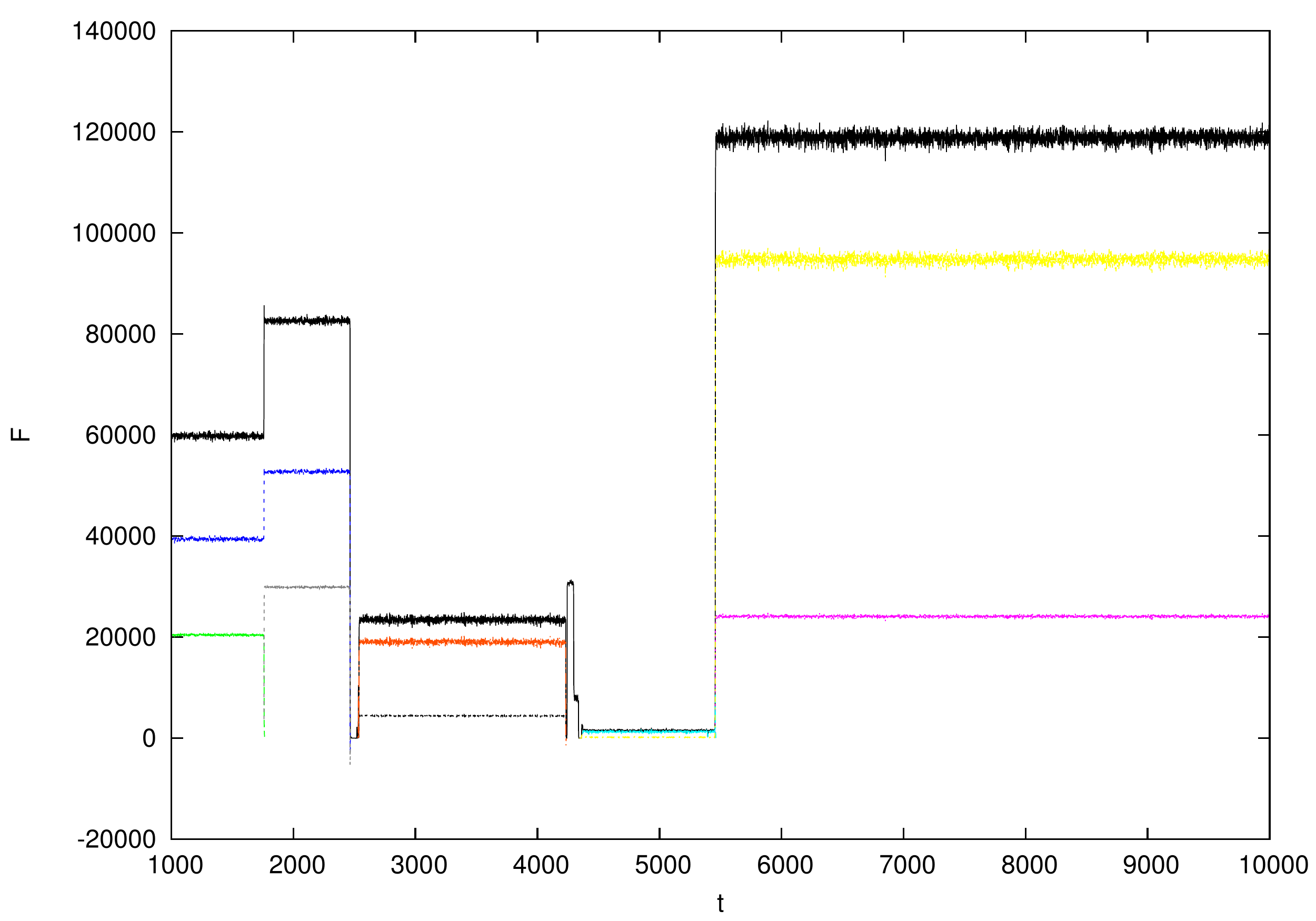}
 \captionof{figure}{
 Fitness versus time for 10000 generations of $\beta = 1, \gamma = 0$. 
 The uppermost black line is the
 total fitness $F$, the coloured lines are the species fitnesses $F_a$. Different colours
 represent different species. \newline
 }
\includegraphics[angle=-90,width=0.45\textwidth]{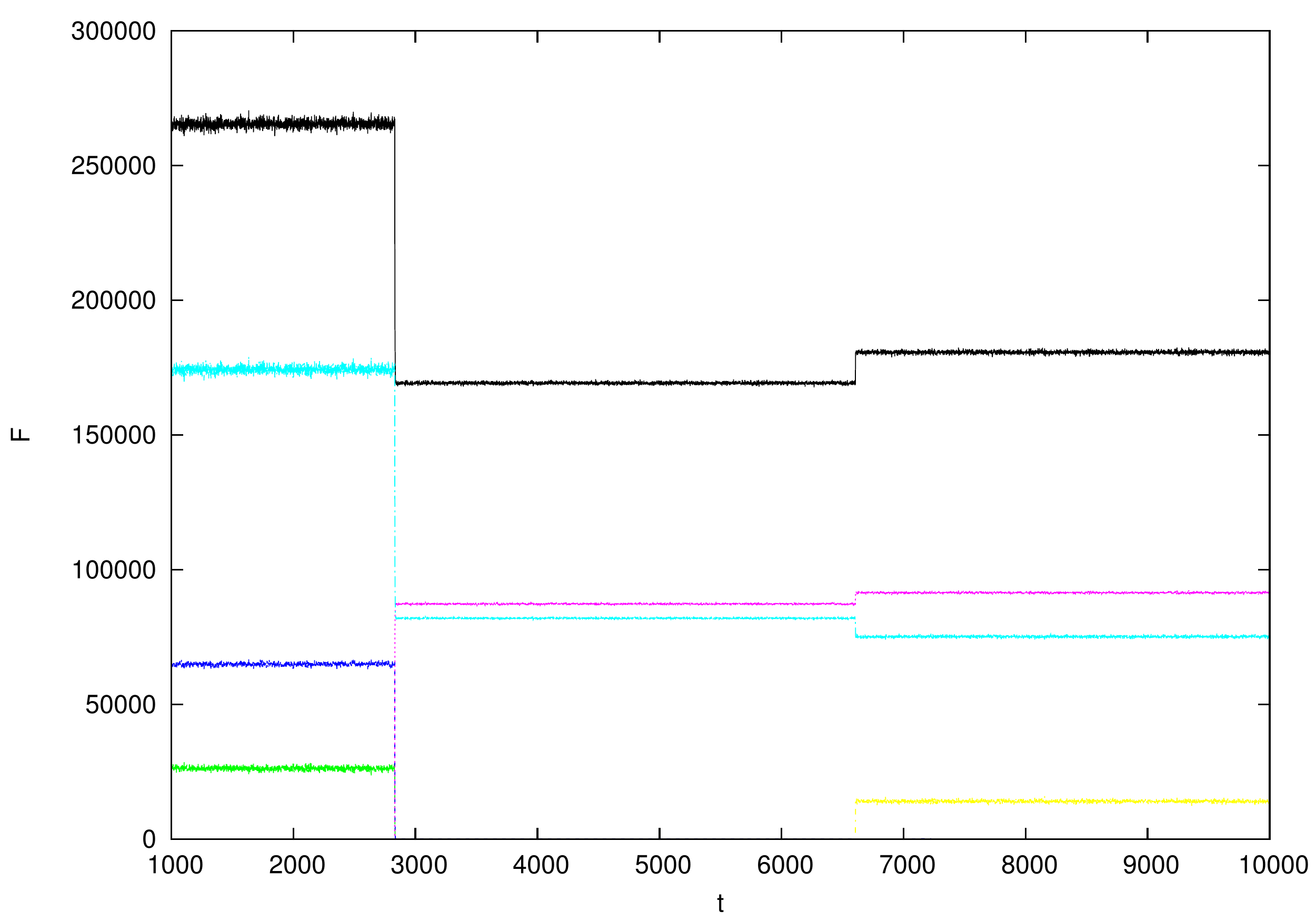}
 \captionof{figure}{
 Fitness versus time for 10000 generations of $\beta = 0, \gamma = 1$. 
 The uppermost black line is the
 total fitness $F$, the coloured lines are the species fitnesses $F_a$. Different colours
 represent different species.\newline
 }

We show in figures 1 and 2 the total fitness of the system and the fitness of the largest
species as functions of time (in generations)
for purely local $(\beta, \gamma) = (1,0)$ and purely global $(\beta, \gamma) = (0,1)$ decisions. 
In either case the system's fitness can go
{\it up and down} as we attempt to optimise. 
Typically the system is at a 
pseudo-equilibrium where all large \emph{core} species have mutually positive interactions.
The large moves (quakes) between different pseudo-equilibria happen
very quickly and are caused by two processes. 
The first is the arrival of parasite species, $p$, which has strong
asymmetric interactions with the other species $J_{pa} > 0, J_{ap} < 0$. These species usually destroy the equilibrium
and allow a completely new one to arise. The second process is the arrival of a species, $n$, having positive interactions
with everyone $J_{na} > 0$, causing rearrangement of the relative populations.
The situation is qualitatively similar to what has been observed in 
the TNM~\cite{Becker:2013}

Note that that the system is not static! Even after many generations quakes occur.
Different mechanisms allow exploration even at a local equilibrium. One mechanism is via rare
long-jumps, when there are many mutations. Sometimes these will produce destabilizing mutants, though this
becomes rarer with time. Another process is neutral mutation - 
often there are species present at a low population level which don't interact with
any of the other species. These can mutate into other neutral species and in this way, without
changing the fitness, the landscapes can be explored. 
Also crucial is the fact that fitness does not always increase after a quake. As we will show, it is true
that the average movement is towards higher system fitness, but progress can be complicated. 

It may seem counter-intuitive that the system's fitness can decrease even with
global decision making. This happens because we can only forecast the effect of small changes
in the short term. For example, it may increase the fitness of 
a system to introduce a single member of a new species, $b$, that has positive interactions with all the other 
extant species except $a$. As $b$ grows it may kill off $a$ and 
this can cause the interaction network to collapse. This mechanism can be see in
figure 2 at around the $3000^{\text{th}}$ generation.

\subsection{Averages}
\includegraphics*[angle=-90,width=0.42\textwidth]{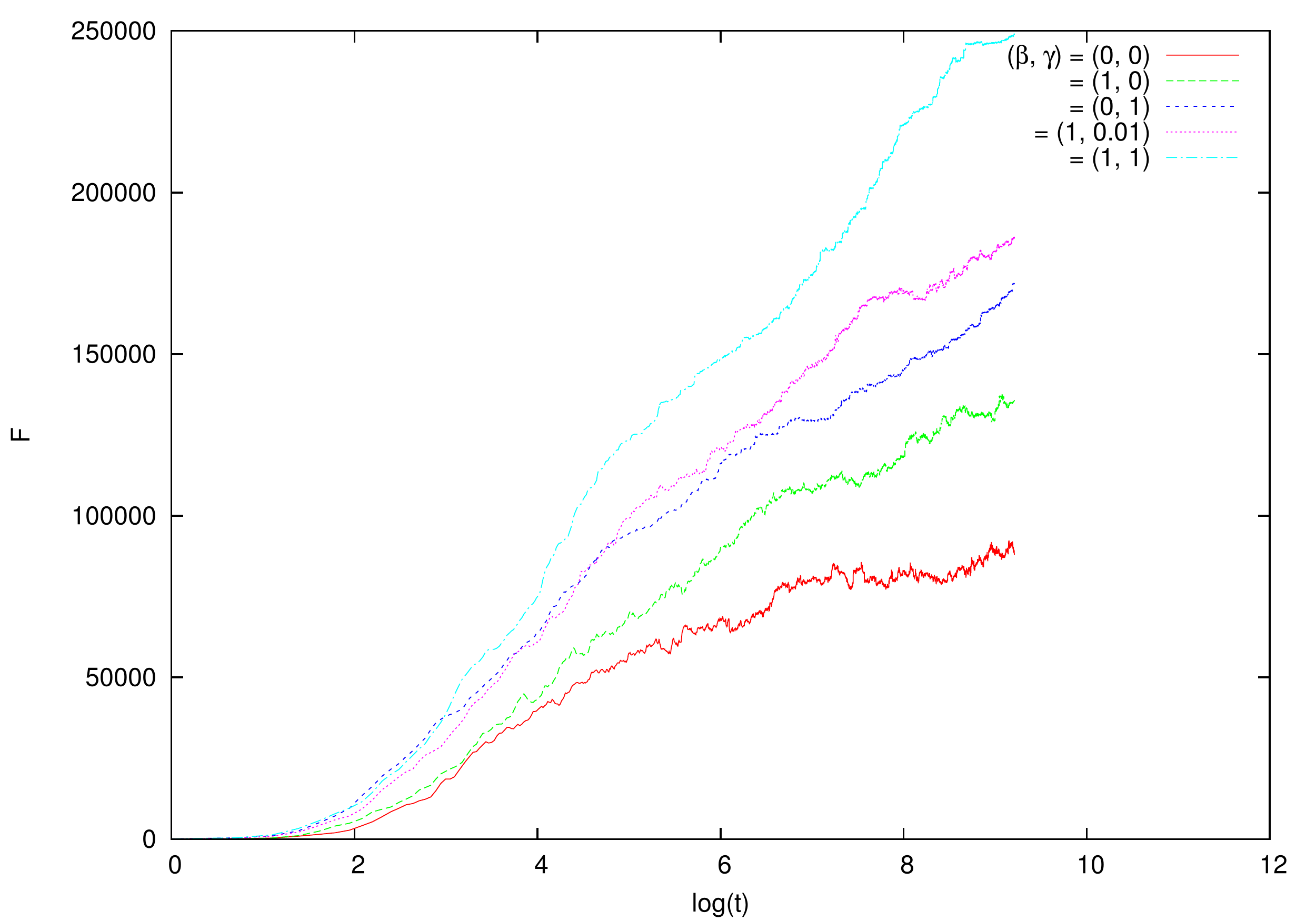}
 \captionof{figure}{
 Fitness versus $\log$ time averaged over 400 runs. 
 }
 \includegraphics*[angle=-90,width=0.42\textwidth]{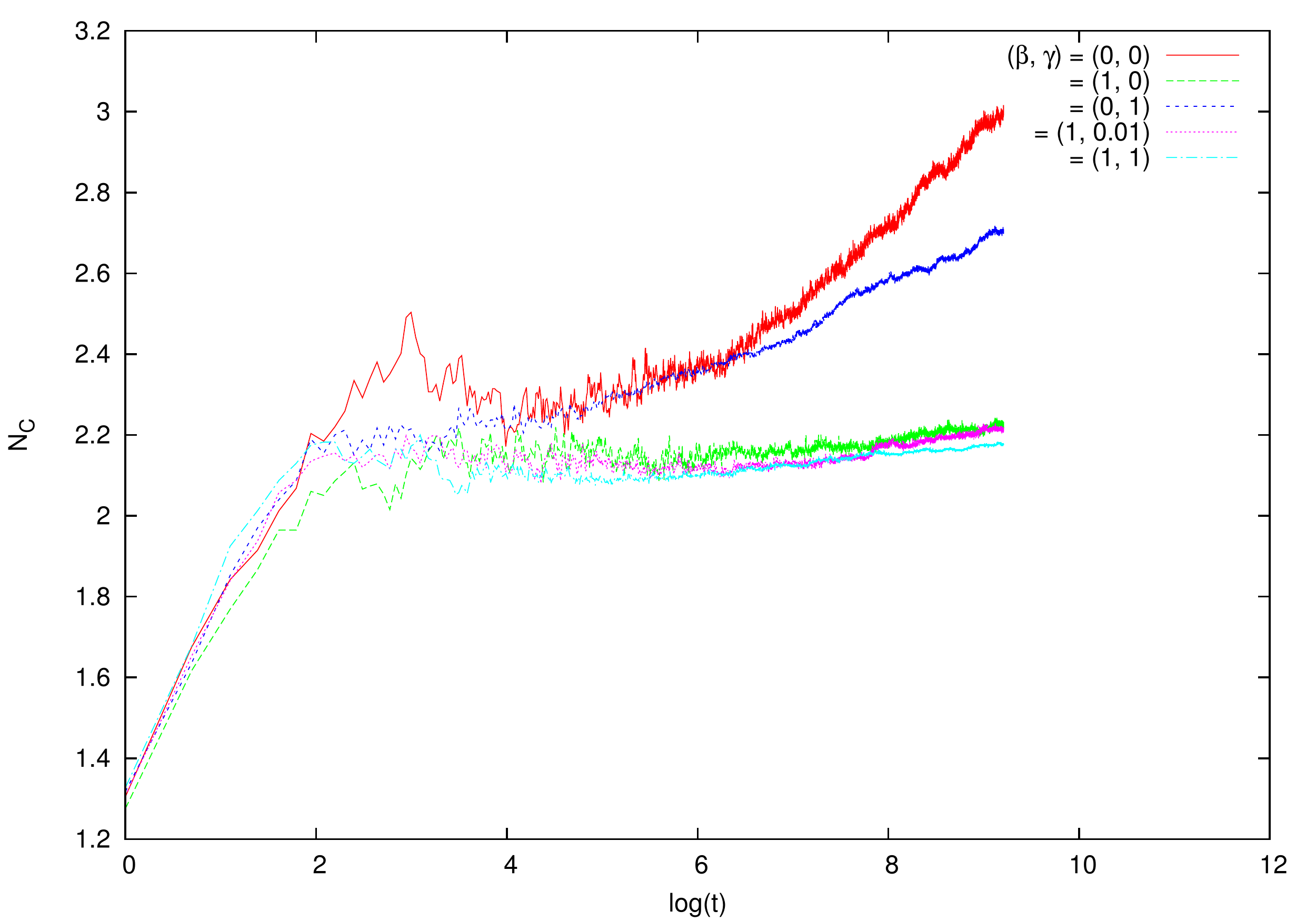}
 \captionof{figure}{
 $N_C$ versus $\log$ time averaged over 400 runs. \newline 
 }
Figure 3 is the average system fitness (over 400 runs) with $(\beta, \gamma) = \{(0,0), (0,1), (1,0), (1,0.01), (1,1)\}$.
These are respectively irrational, local, global and mixtures of local and global descision making.
Fitness increases on average in all cases, with the mixture of global and local decisions
producing the fastest increase. Figure 4 shows the number of core species $N_C$.
A species is in the `core' if its population is greater that 5\% of the population of the
most populous species and otherwise it is in the `cloud' \cite{Becker:2013}. Irrational 
decisions and global decisions produce large cores while 
adding local decision making drastically reduces core size. Compared to the irrational model (TNM) the clouds are small 
- less capacity is spent generating and supporting unfit mutants in the rational models leading to lower diversity.

The pair $(\beta, \gamma) = (1, 0.01)$ is interesting. In practice evaluating global fitness
will be extremly difficult, the uncertainty will be greater than
for local descisions as we require much more information. We model this, very schematically, by putting $\gamma \ll \beta$. 
Even with quite a large ratio, $\frac{\beta}{\gamma} = 100$, there is still something
to be gained for the system, and hence the typical species, by attempting to globally optimise.

\section{Discussion}
The TDM and the TNM are quite similar in their qualitative features.
The core-cloud structure of the TNM remains, though with much smaller clouds as fewer obviously unfit mutants are generated.
The same parasitic and complementary process that disrupt the TNM pseudo-equilibria occur in the TDM, however
neutral mutations are much more important in the TDM as a source of diversity. In the TNM individual agents
have a natural interpretation as individuals of a species, in the TDM, as in \cite{us:2016}, we treat the agents
simply as a way to evolve the system and measure the size of a species. We also note here that the TNM, and therefore
the TDM, have a strong similarity to a glassy system and evolve by crossing successive `entropic barriers' 
\cite{Becker:2013, Andersen:2016}, which in the TNM/TDM results in the gradual increase of core size.

Searches on the system landscape by locally optimising agents
proceed uphill on average, but not without frequent fluctuations. A global perspective
leads to faster fitness growth though in practice trying to forsee the effect of 
a decision on the whole economy is difficult. `Selfish' local-optimisation requires less information and will be easier in practice. 
Our results suggest that even adding relatively
inaccurate global forecasts to accurate local ones is beneficial for the system and hence for the
average company. However the `tragedy of the commons' means that, without enforcement, this environment
will be vunerable to exploitation by selfish agents.

It would be a leap to extrapolate into management or policy suggestions. Instead we want to focus
on improving our mental model of descision making in complex systems. The fitness landscapes 
of Sewall-Wright, Kauffmann and the NK-model encourage thinking about decisions in isolation.
The lack of interaction leads to the dangerous misconception that what is good for one
agent is always good for the whole system. Fitness is not absolute, it is relative.
I can be brilliantly placed, until a new product comes along and wipes out my
main customer. We hope to have convinced our readers that interacting landscapes and
the encompassing system landscape are better pictures of what is happening in complex sociological and economic systems. 
We also hope that the TNM, TDM and variants can find wider applicability in this domain, as an evolution 
of the NK-model into an interacting, multi-species modelling framework.

\section{Appendix: Simulation Details}\label{sec:app}

Our implementation of the TNM has been described in detail in \cite{us:2016} and we follow most of the
same steps. We use genomes of length
$L=20$ and set the carrying capacity parameter $\mu = 0.1$. We set the elements of the interaction matrix
$J_{ab}$ to zero with a $75\%$ probability and choose values for the rest by multiplying two independent 
gaussian random numbers together (mean $0$ standard deviation $1$) for each non-zero pair. This 
is to avoid having to store the matrix $J$ with $2^L \times 2^L$ entries. We then multiply
the non-zero elements by a factor $1/C = 100$. The death probability is $p_{kill} = 0.2$ and the mutation
probability is $p_{mut} = 0.01$. We initially seed the system with $500$ members of a randomly chosen species.

\end{multicols}

\end{document}